\documentclass[fleqn,10pt]{wlscirep}
\usepackage[utf8]{inputenc}
\usepackage[T1]{fontenc}

\usepackage[
    natbib=true,
	style=numeric,
	citestyle=numeric,
    sorting=none
]{biblatex}
\usepackage{lineno,hyperref}
\usepackage{braket,amsfonts}
\usepackage{amsmath, amsthm}
\usepackage{array}

\usepackage[caption=false]{subfig}
\usepackage{algorithm}% http://ctan.org/pkg/algorithms
\usepackage{algpseudocode}% http://ctan.org/pkg/algorithmicx
\newtheorem{theorem}{Theorem}
\newtheorem{defn}[theorem]{Definition} 
\newcommand{\vbf}[1]{\boldsymbol{\mathbf{#1}}}

\title{Agent-Based Modelling: An Overview with Application to Disease Dynamics}

\author[1,*]{Affan Shoukat}
\author[1]{Seyed Moghadas}
\affil[1]{Agent-Based Modelling Laboratory, York University, Toronto, ON M3J 1P3, Canada}
\affil[*]{ashoukat@yorku.ca}
\begin{abstract}
    Modelling and computational methods have been essential in advancing quantitative science, especially in the past two decades with the availability of vast amount of complex, voluminous, and heterogeneous data. In particular, there has been a surge of interest in agent-based modelling, largely due to its capabilities to exploit such data and make significant projections. However, any well-established quantitative method relies on theoretical frameworks for both construction and analysis. While the computational aspects of agent-based modelling have been detailed in existing literature, the underlying theoretical basis has rarely been used in its construction. In this exposition, we provide an overview of the theoretical foundation of agent-based modelling and establish a relationship with its computational implementation. In addition to detailing the main characteristics of this computational methodology, we illustrate its application to simulating the spread of an infectious disease in a simple, dynamical process. As the use of agent-based models expands to various disciplines, our review highlights the need for directed research efforts to develop theoretical methods and analytical tools for the analysis of such models.
\end{abstract}
\begin{document}

\flushbottom
\maketitle
\thispagestyle{empty}

% \noindent Please note: Abbreviations should be introduced at the first mention in the main text – no abbreviations lists. Suggested structure of main text (not enforced) is provided below.

\section{Introduction}
\label{sec:intro}

The seminal work of the American economist Thomas Schelling in 1971 \cite{schelling1971JMathSociol} showed that computational and simulation approaches can be applied to understanding the universal principles of any `complex adaptive system'. While there is no single definition of a complex adaptive system, it is generally accepted that such a system has many individual parts working together to generate the system behaviour referred to as macro dynamics. Complex adaptive systems are common in both nature and society. For example, the immune system is a highly advanced biological system comprised of individual cells, chemicals, and signals  working together to create a complex network. Although it is not feasible to capture this complexity in its entirety, many key characteristics of the underlying system can be reconstructed \textit{in-silico} (i.e., computer) environment with sufficiently detailed programming of the elementary mechanisms. Schelling's work laid the foundation for a new field of research on socio-economic systems, in which the natural unit of decomposition is the `individual' rather than the observable or equations, often termed as individual-based modelling. With the rapid advances in computational theory, this field has led to the evolution of \textit{Agent-Based Modelling} (ABM) computational systems.

Today, methodologies studying complex adaptive systems in a qualitative sense have shifted to systematically investigate them by using ABM through a disaggregation of the systems into their individual components that have their own characteristics and behaviours \cite{bonabeau2002ProcNatlAcadSci, macal2010JSimul}. 
An ABM computational system is a dynamical model that consists of a collection of abstract objects (i.e., agents) embedded in an \textit{in-silico} environment and interact through a set of prescribed rules. This type of model is often implemented computationally by using deterministic input-output functions, typically coded in a structured or object-oriented programming language. 
The `agents' in ABM represent the individual components of the complex system under study. Each agent individually \textit{perceives} its situation, makes \textit{decisions}, and performs \textit{actions} according to specific rules. These rules can be simple or complex, deterministic or stochastic, and fixed or adaptive. Thus an ABM encodes in a computer program a set of rules that describe the interdependencies between the individual components as the system evolves in time. While often the formal set of rules are simple, large scale models can incorporate neural networks, genetic algorithms, and other machine learning techniques for realistic agent's behaviour and adaptation \cite{vanderhoog2017ArXiv170606302Q-Fin}. Since each agent is modelled individually, there is no central controlling agency nor explicit language that describes the global dynamics of the system. As a consequence, ABM allows for an investigation into the universal properties of a complex system, including: \textit{heterogeneity} since agents can be modelled individually, \textit{adaptation} since the model is dynamic, \textit{space and scale} since an arbitrary number of agents can be embedded in this virtual environment, and \textit{non-linearity} since the model can track individual agents separately.
 
An agent-based model has three main advantages: (i) it can capture the global dynamics as a result of the local interactions between heterogeneous agents; (ii) it allows for the construction of a system in the absence of knowledge about global interdependencies between individual components; and (iii) it provides the flexibility required to study the system's complexity in comparison to traditional equation-based aggregate-level modelling strategies. Given these advantages, and a growing body of literature with the use of ABM in various disciplines (e.g., economics \cite{chen2012KnowlEngRev,north2010Complexity,cristelli2011ArXiv11011847PhysQ-Fin}, ecology \cite{deangelis2014F1000PrimeRep}, sociology \cite{macy2002AnnuRevSociol}, geography \cite{brown2005IntJGeogrInfSci}, finance \cite{brock1998JEconDynControl}, military \cite{ilachinski1997,moffat2006JApplMathDecisSci,hill2004JDefModelSimulApplMethodolTechnol}, and healthcare \cite{effken2012CINComputInformNurs}), a better understanding of this computational system is essential for advancing quantitative science.  

In this exposition, we provide an overview of ABM by describing its fundamental components. We detail how an agent-based model is computationally constructed, validated, calibrated and executed. This is followed by a mathematical formalism of ABM which is founded upon partial recursive functions and well-established Markov Chain processes. Finally, we present an application of ABM in a simple model of disease transmission dynamics to illustrate how the theory and computation are integrated.

\section{Emergent global dynamics in ABM}
\label{sec:EGD-ABM}
Traditional models of complex systems are typically formulated using mathematical equations. Examples include dynamical systems such as the Lotka-Volterra equations describing predator-prey interactions \cite{jopp2010modelling}, and the \textit{Susceptible-Infected-Recovered} models of disease epidemics \cite{kermack1927ProcRSocMathPhysEngSci}. Despite their usefulness that have been exemplified in the literature, these models often have significant limitations that arise from the treatment of all or groups of individual components of the systems as largely homogeneous entities, i.e., a \textit{representative class}. For example, in differential equation models of disease transmission, all individuals in a sub-population are represented with the same characteristics. On the other hand, ABM enables the generation of system dynamics in a `bottom-up approach' in which a single homogeneous model is replaced with a population of individual models, each of which is an autonomous decision-maker (i.e., the agent). This heterogeneity includes not just the variation in individual agents, but also the interactions and network topology. Running such a model simply amounts to instantiating an agent population with initial conditions and iteratively letting the agents interact by executing a set of rules that define them. Of course, if the system is stochastic in nature, then capturing randomness is of particular importance. As a result of heterogeneities and stochasticity, enigmatic global dynamics including fixed points, cycles, dynamic patterns, and long transients emerge from the local properties and interactions among agents. These emergent macro dynamics, although not explicitly programmed in the model, can have properties that are decoupled from those of the individual components. The emphasis on modelling the heterogeneity of agents and the emergence of global behaviour from local interactions is a distinguishing feature of ABM. 

\section{Absence of global interdependencies}
\label{sec:AGI}
Agent-based models are particularly suited to complex systems in which the dynamics of its constituent components are more understood than the overall dynamics of the system. They are ideal for modelling systems in which agents' behaviour is non-linear and heterogeneous, and includes learning and adaptation, temporal and spatial correlations, and non-Markovian properties \cite{richiardi2018Agent-basedModelsb}. In this sense, contrary to what the term \textit{agent} suggests, the concept of an agent enables us to represent any physical object that can be programmed, provided we have a clear understanding of the object.  Agents can represent particles, cells, humans, various species, and spatial entities such as buildings and organizations. Consequently ABM can, in principle, incorporate any complex behaviour that can be observed experimentally provided we have a qualitative description of the underlying mechanisms and components. Of course, in practice agent-based models are limited by finite computational resources, time investment, and incomplete knowledge, and therefore will inevitably demand a balance between the desired complexity of a system and available resources.

\section{Rigour in ABM}
\label{sec:FABM}
A drawback to the ABM approach is the lack of a well-established theory, which can be used for rigorous analysis similar to equation-based models.  Analytical and differential equation dynamical systems, provide a formal framework for the organization and analysis of knowledge and theoretical results. For example, such models are equipped with various tools for stability and bifurcation analyses, and even sensitivity analysis of their outputs, which allows them to communicate mathematically without ambiguity. Agent-based models, on the other hand, often do not make use of any explicit mathematical equations but exploit computational simulations, implemented in a programming language to elucidate the complex dynamics underpinning the system. However, the idea that the lack of formal mathematical tools prevents any sort of formal analysis in ABM is misguided. Indeed ABM computational systems, by virtue of being computer programs, can and do utilize a well-defined set of functions which relate inputs to outputs, in either a deterministic or stochastic fashion, and unambiguously define the global dynamics and any eventual equilibria of the system \cite{epstein2006HandbookofComputationalEconomics, delligatti2018,richiardi2018Agent-basedModels,laubenbacher2007ArXivPreprArXiv08010249}. For example, a proposed mathematical representation of an agent-based model \cite{laubenbacher2007ArXivPreprArXiv08010249} is a \textit{discrete-time dynamical system} over a finite set of states. In this framework, the state of the model is fully specified by a vector taking values over a finite field $\mathbb{K}$. An update function transfers a given state into another state based on rules of the complex system. The model dynamics are generated by repeated iteration of this function. Another approach \cite{veliz-cuba2010Bioinformatics} is to derive a polynomial dynamical system, where the input-output functions are defined by polynomials, which makes it amenable to powerful symbolic computational capabilities. This allows for the computation of equilibria and analysis of the model to symbolically solve a system of polynomial equations. 

From a formal point of view, agent-based models are also `Markov Chains' \cite{banisch2012ArXiv12093902NlinPhysicsphysics, banisch2011ArXiv11081716Phys, IzquierdoJass}, which provide a rigorous mathematical basis of ABM by linking the micro-description of the system to the complex local behaviours. These chains establish how the corresponding global dynamics of an agent-based model are obtained by a projection construction, and how the model's long-term properties are given by the ergodic theorem of Markov processes. Analysing an agent-based model as a Markov Chain can make apparent transient dynamics, asymptotic behaviour, and stochasticity that were otherwise not evident.  More generally, ABM can be naturally classified as hidden models that relate a set of observable variables to a set of latent variables \cite{lux2018JEconDynControl}. In addition to formal mathematical frameworks, there is an established protocol for an agent-based model specification developed by \cite{grimm2006EcolModel}. This protocol, the so-called `ODD' (Overview, Design concepts, Details), describes a standard template for the model construction, reproducibility, and update functions.

Another common objection to ABM is that a single realisation of the model under the condition of randomness cannot be used to derive any reliable conclusion of the results. This is partially addressed in \cite{Axtell20001WA} where the computer programs are provided as sufficiency theorems approach of Newell and Simon. 
That is, when an agent-based model, call it $A$, produces a result $R$, it establishes a sufficiency theorem which is the formal statement `if $A$ then $R$'.  In other words, each run of the model is a logical theorem that reads `the output of an agent-based model follows with logical necessity from applying to the input a formal set of rules that define the model'. 

Despite these criticisms, the creation of ABM in a bottom-up approach offers several advantages over conventional equation-based models, since the output of ABM tends to be more visual, pattern-oriented, and can range from micro to macro levels of the system. These advantages arise from two fundamental differences between agent-based and equation-based models. First, equation-based models express the relationships among observables and their evolution over time as input of the system. ABM, on the other hand, begins not with equations that relate observables to one another, but with a set of rules which govern the agents' behaviour and their interactions in the environment, as described in Section \ref{sec:SCs}. In this context, direct relationships among the observables become an output of the model. Second, equation-based models tend to focus on the observables at the system (macro) level, which often simplifies their formulation using closed-form equations. Hence, they may become mathematically tractable for a thorough analysis using established theories of dynamical systems. However, many observables are measurable characteristics at the micro level and may differ from those aggregated at the system level. Since ABM relies on emergent dynamics from interacting agents and environment, capturing these observables at the micro-level becomes possible.

\section{Structural Components}
\label{sec:SCs}
\begin{figure}[htb!]
	\centering
	\includegraphics[width=\textwidth]{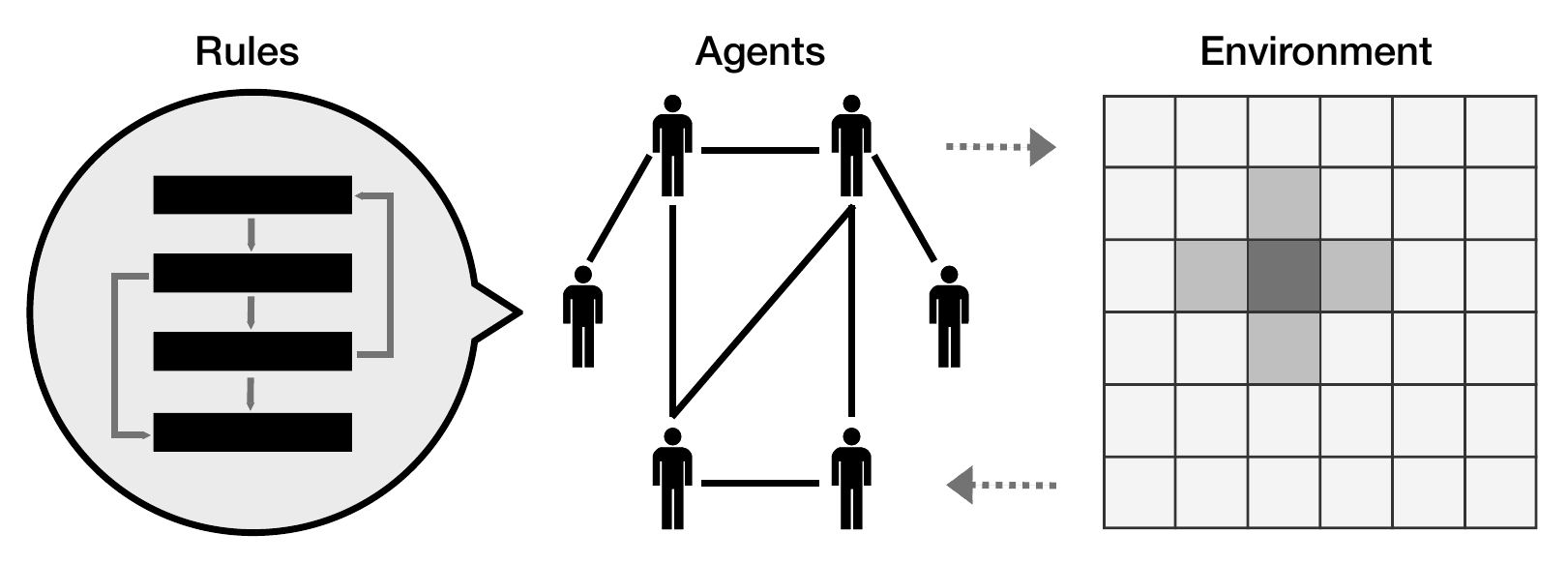}  
	\caption[The general structure of an ABM computational system.]{The general structure of an ABM computational system. Agents are characterized with a set of properties, connected and interact through a lattice environment system.}  
	\label{fig:overallstructure}
\end{figure}

Any agent-based model has three fundamental components:
\begin{itemize}
	\item[{1)}] A set of agents that represent the smallest components of the system being modelled.
	Since every agent-based model is a computer program, agents are often implemented as virtual computational \textit{objects}. In computer science, an object is a data structure consisting of variables, functions, or methods, and is a value in memory which is referenced by an identifier. There is a large degree of similarity between a computational agent and the concept of an object (or structure) in a programming language. An agent-based model can be seen as a set of objects that share the same properties and the same rules. The concept of an agent as being a computational object makes it clear that this basic unit is abstract.  An agent is made concrete by translating sufficient properties of the real-world component into a suitable formulation in the programming language.   
	\item[{2)}] An \textit{in-silico} environment that allows agents to change their spatial and relational associations. In most complex systems, the concept of physical space or spatial network is significant to the global dynamics. This concept is difficult to model in traditional analytic models. However, in ABM, it is rather simple to have the agent interactions mediated by a virtual space.  
	\item[{3)}] A set of rules that define the level of connectedness and modes of interaction between agents. Each agent can be assigned a unique set of rules or a single one that can be applied to all (or a group of) agents.
\end{itemize}
To integrate these three components, an agent-based model requires a computational framework or a \textit{simulator engine}. This simulator is responsible for driving the ABM by repeatedly (i.e., by iteration) executing the rules that define the  agents' behaviours and interactions. This iterative process often operates over a time-step or discrete-event simulation structures. In the course of these iterations, the simulator also calculates the aggregate results of the model which can be re-injected back into the evolving behaviour of the agents.  The structure of a typical agent-based model is shown in Figure~\ref{fig:overallstructure}. 
 
\subsection{Agent Structure and Properties}
\label{sec:agents}
Since an agent-based model is an abstraction of a real-world phenomenon, an agent ideally represents a component of the complex system being reconstructed.  When using ABM approach, one needs to systematically recognize which components of the system can be translated into agents. Then, a decision will need to be made on the level of abstraction and the details that may be included for each agent.  An acceptable compromise between realism and simplicity is required. If the level of abstraction is too low, the model may fail to faithfully capture the system dynamics. On the other hand, a one-to-one mapping of the real-world component to an agent is likely unnecessary, impossible, and/or computationally overwhelming. One will need to have valid hypothesis on the underlying processes or fundamental mechanisms that are sought to be explained. Ideally, the level of abstraction is justified by utilizing empirical data as well as expert opinions. Previous work \cite{doran2006Agent-BasedComputationalModelling} suggests two principles that should be followed for implementing a set of rules: (i) agents should be as abstract as possible subject to the requirements that any rule attributed to them must either be reliably derived from empirical observation or be subject to experimental observations, and (ii) assumptions based on pre-conceptions are avoided \cite{doran2006Agent-BasedComputationalModelling}. 
Several other studies \cite{helbing2012,richiardi2012KnowlEngRev, macal2010JSimul, crooks2012Agent-BasedModelsofGeographicalSystems,epstein2006HandbookofComputationalEconomics,grimm2006EcolModel,laskowski2014Biomat2013} have also identified similar key characteristics: 
\begin{itemize}
	\item \textit{Self-contained and autonomous}. Autonomy implies that there is no central authority that controls the agents' behaviour. An agent can be thought of a model in itself, capable of processing information and making decisions.  Agents are free to interact among each other and move in the virtual environment they reside in. 
	\item \textit{Heterogeneity}. Each agent represents a unique component of the complex system.
	For example, an agent representing a human in some population model can have age, sex, health status, and location as dynamic attributes. 
	\item \textit{Active}. Agents can be `goal-oriented' \cite{laskowski2014Biomat2013}, that is they try to achieve a goal but not necessarily maximize utility or bounded rational \cite{secchi2017TeamPerformManagIntJ} in which agents are generally assumed to be rational optimisers with complete access to information and possibly restricted analytical ability (through heterogeneity). 
	\item \textit{Internal state}. Agents have an `internal state' (i.e., data represented by variables) and can communicate this state to the model by message-passing or signal protocols. 
\end{itemize}

Once the agents' characteristics are defined, the next step is to implement a formal set of functions (or the rule-set) that define the `perception-decision-action' (PDA) cycle for every agent \cite{laskowski2014Biomat2013, drogoul2009Multi-AgentSystems}. An agent-based model systematically and iteratively gives each agent the chance to perform a PDA cycle. Mathematically speaking, let $x_a$ represent the internal state of an agent $a$, which is essentially a list of quantitative variables and internal parameters associated with the agent's current situation, i.e., a vector taking values over some finite field $\mathbb{K}$ (often $\mathbb{K} = \mathbb{R}^n$) describing the state of the agent at a given time.  Let $\vbf{x} \in \vbf{X} = \{ x_a \}_{a \in \Lambda}$ denote the global state of the model, where $\Lambda$ is the set of all agents. Then, with notation borrowed from \cite{drogoul2009Multi-AgentSystems}, the PDA cycle of the agent $a$ is a set of functions that: 
\begin{itemize}
	\item  $\textsc{Perception}()_a : \vbf{X} \longrightarrow \mathcal{P}$ which computes a \textit{percept} $p \in \mathcal{P}$ using the global state of the model $\vbf{x}$.  Intuitively, the function processes this state, and returns data such as the coordinates of nearby agents or objects, or possible locations for a roaming agent. Thus, $p$ is often an $n$-tuple consisting of quantitative and categorical information.
	\item $\textsc{Decision}()_a : \mathcal{P} \rightarrow \mathcal{D}$ which is a core function executing the rules of the agent's behaviour given their \textsc{Perception}. Decision functions can be arbitrarily complex, encapsulating the bulk of the model's logic, ranging from simplistic fuzzy rules to complex behaviours modelled by neural networks, logic systems, artificial intelligence, or a hybrid multi-layer system. These functions are often imputed by using empirical data and expert opinion.  
	\item  $\textsc{Action}()_a : \mathcal{D} \times \vbf{X}  \rightarrow \vbf{X}$ computes the new internal state $x_a$ of the agent by acting on a decision $d \in \mathcal{D}$. Actions can be in the form of, for example, discrete messages, or signals that are transferred between agents. In response to a message, an agent may change their internal state, modify the environment, or respond back with another message, but should not be able to modify the internal state of other agents. 
\end{itemize}
\begin{figure}
	\includegraphics[width=\textwidth]{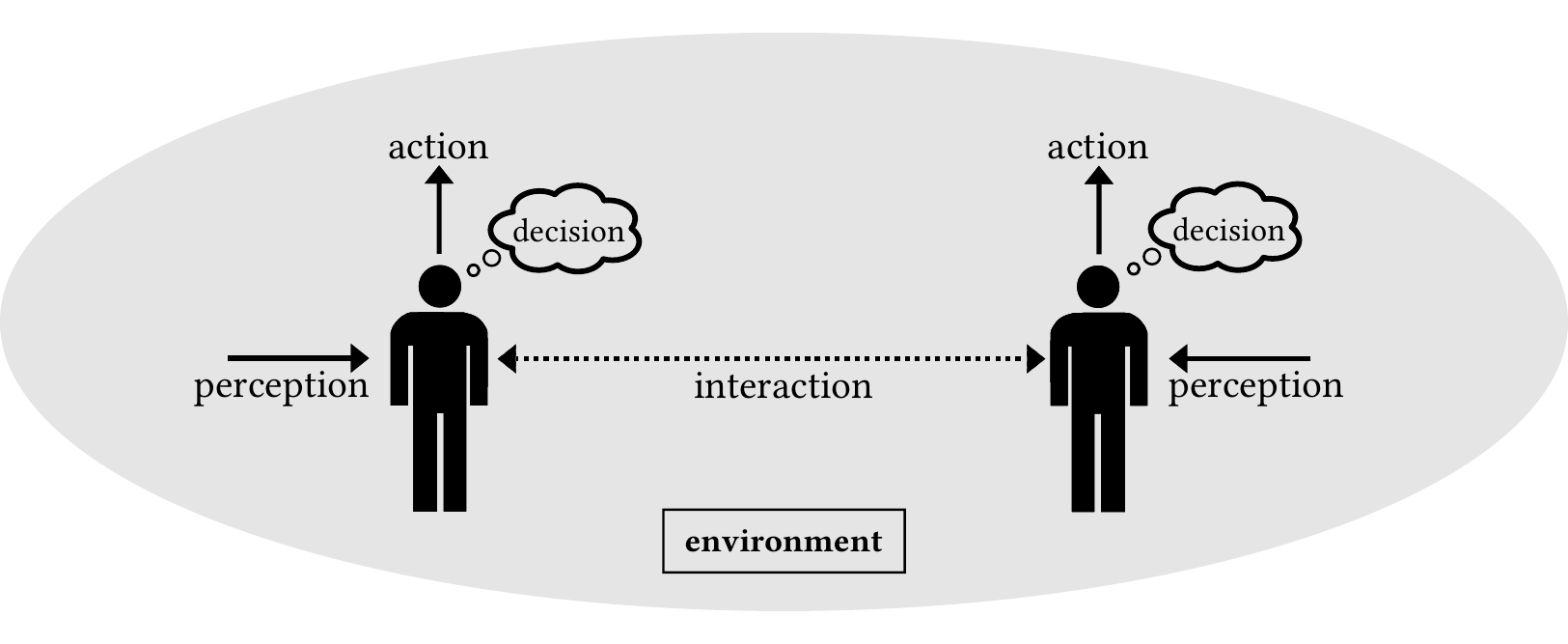}
	\caption[The \textit{Perception-Action-Decision} cycle of an agent.]{A visual representation of a PDA cycle which formally implements the rule-set to guide an agent's behaviors and interactions.}
	\label{abm:fig:pdacycle}
\end{figure}

\subsection{The Virtual Environment}
\label{sec:environment}
The second fundamental component of an agent-based model is the \textit{in-silico} environment, which represents the physical space of the corresponding real-world system. The environment defines the spatial associations of an agent and the conditions for the PDA cycle to carry out, i.e., the environment is part of the input state to the \textsc{Perception} function of every agent. The environment may also include passive objects such as roads and buildings or resources such as wealth and healthcare. As agents move in the environment, their location can be tracked by a dynamic variable. Agents may also be spatially implicit, meaning that their location within the environment is irrelevant.

\begin{figure}[ht!]
	\includegraphics{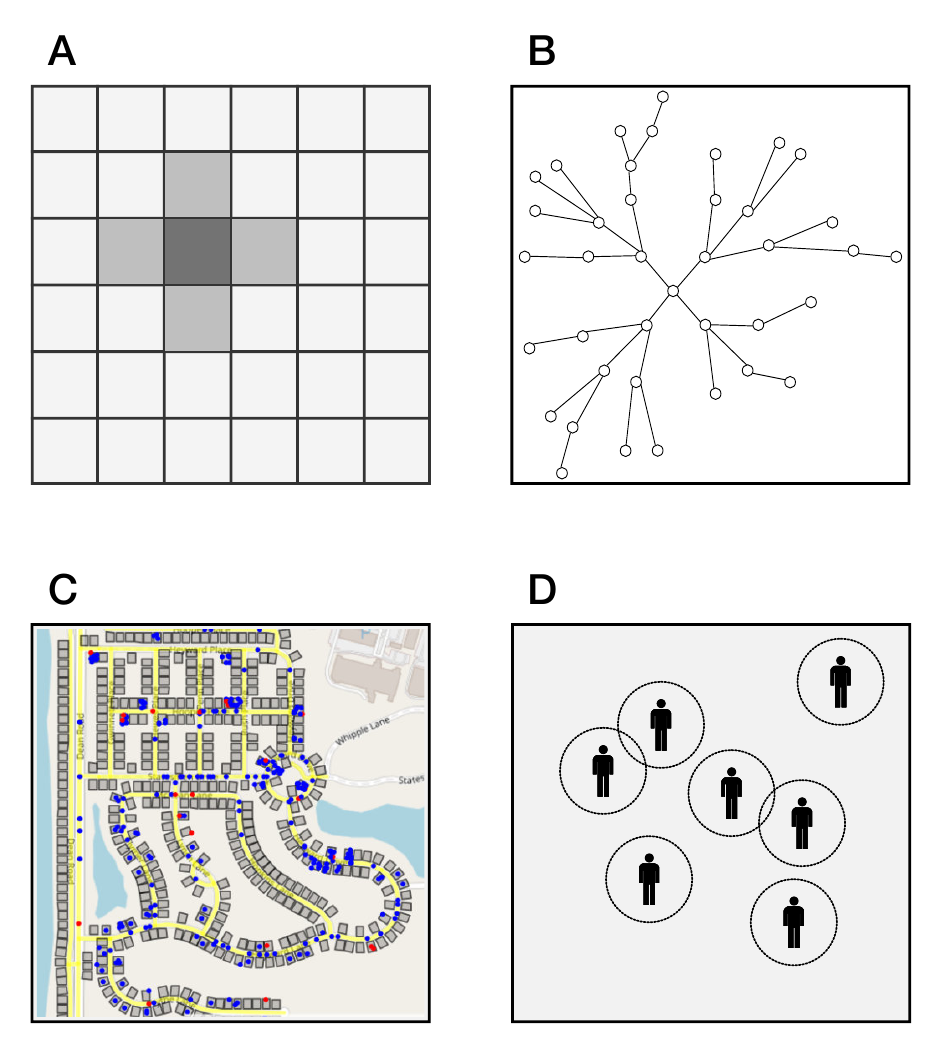}
	\caption[Various types of environment topologies used in ABM.]{Different environment topologies for agent interactions and movement: (A) cellular automata (von Neumann neighbourhood), (B) network relational topology, (C) geographic information system, and (D) free-roaming with neighbourhoods.}
	\label{fig:environments}
\end{figure}

Modelling the environment is often via a `discrete topology' of connected and bounded space units. The topology of the environment defines possible interactions and relationships between agents. There are two main types of spatial environments:
\begin{enumerate}
	\item[{1)}] The most common type is the discretized environment consisting of a finite grid of cells in one or more dimensions with integer coordinates \cite{drogoul2009Multi-AgentSystems}, which provides a simple representation of physical space, e.g. GIS based environments. This type of environment provides an easy mechanism for agents to interact with others who share similar coordinates or reside in neighbouring cells. 
	The simplest form of spatial environments are described by cellular automata models where the environment consists of a square lattice, divided uniformly into `cells'. However, in cellular automata, the cells are interpreted as agents and there is no distinction between the agents and the cells that create the lattice environment. ABMs extend this topology by decoupling the agents from their cells. Agents can then move from one cell to another and interact with different parts of the grid, resulting in a set of neighbours that constantly change as the simulation proceeds. 
	\item[{2)}] The second type is a `relational environment' in which a link between agents $a_1$ and $a_2$ defines a network topology. Relational environments are often defined by graph- or node-based constructions and there is no distinction between agents and the nodes of the network. Both directed and undirected graphs can be supported and can be static or dynamic \cite{macal2010JSimul}. In static networks, links are fixed and do not change. In dynamic networks, links and nodes are determined endogenously according to the mechanisms programmed in the model.  
\end{enumerate}
Figure~\ref{fig:environments} shows examples of different environment topologies. 
Agents typically interact with a subset of all other agents, which are located in the agent's `neighbourhood'. 
In complex and large-scale agent-based models, spatial and relational environments can be intertwined to offer a more granular approach. The environment may further have its own set of properties relevant to the real-world systems. For example, in modelling disease-transmission at the age-group level, it might be relevant to identify sections of the environment as School, Residential, and Business. The environment may also respond to messages from agents or change with time (i.e. a dynamic environment), and can even generate or delete agents.

\subsection{Rules}
\label{sec:interaction}
Recall that the \textsc{Decision} and \textsc{Action} functions of every agent implement the rules that will affect their internal state and interactions with other agents in a dynamic (possibly changing) environment. The rules are typically derived from published literature, expert opinion, or empirical data. In general, rules are based upon simple \textit{if-else} statements with agents carrying out an action once the specified conditions have been met, but can also be applied using more sophisticated methods like machine-learning techniques and artificial intelligence \cite{wojtusiak2012machine,shoham2007if}. For example, the rules that govern the dynamics of an airborne disease transmission are generally based on interactions between susceptible and infectious individuals in the  environment (e.g., changing their spatial relationship). Because agents are autonomous decision making entities in the model, their decisions to interact determine the possibility of disease transmission. In this process, a simple if-else statement can be executed to check if the interaction is between susceptible and infectious individuals, followed by rejection sampling-based (i.e., Bernoulli) trials to determine the occurrence of disease transmission. At the point of infection if transmission was successful, the timelines for natural history of the disease (e.g., latent, incubation, and infectious periods) for an  infected agent are sampled from the relevant distributions, often derived from observational and epidemiological studies. The execution of such rules and relevant sampling processes induces stochasticity into the system and as a result, different outputs can be produced in each run of the model as emergent phenomena at the system level. Indeed, by utilizing complexity theory and network science, ABM is highly effective in explaining how complex patterns (i.e., a process model)  emerge from micro-level rules (i.e., a pattern model) \cite{wilensky2015introduction} .

\section{Model Implementation}
\label{sec:MI}
Every agent-based model implements a simulator engine that specifies the operating procedures of the agents, drives the PDA cycle, and describes the model evolution in time from an initial condition. The simulator is often implemented in a programming language based on sound principles of computer science \cite{mostacco2011software}. For example, writing independent `functions' (which represent distinct components of the complex system) makes maintenance, debugging, and code reuse easier when considering future changes to the model. Good programming principles also ensures reproducibility of the model across various platforms and operating systems. Frequently used programming languages are Java and C++.

Programming an ABM from the ground up allows for controllability and flexibility of the model. However, this is a time-consuming process and requires expertise and experience in programming. In addition, considerable amount of time may be spent on extraneous activities such as building graphical user interfaces and data collection and input methods. To overcome the constraints posed by time and required expertise, alternative solutions have been developed in simulation platforms that provide a programming environment. Examples of these platforms include NetLogo, RePast, and SWARM \cite{wilensky2015introduction,north2006experiences,minar1996swarm}, which specify pre-defined methods and libraries, enabling novice programmers to rapidly develop and prototype models. However, these platforms have their own set of limitations, including a lack of flexibility if the software does not implement the desired functionality such as incorporating multi-agent complex systems that are high dimensional. Furthermore, a substantial time investment may still be required in learning and utilizing the software.

When a model is implemented, a number of important factors must be taken into account as outlined below.
\subsection{Monte Carlo Simulations}
\label{abm:sec:montecarlosims}
ABMs are generally stochastic in nature and account for randomness found in real-life phenomena. This stochasticity, known as first-order uncertainty, relates to the natural randomness in agent behaviours, interactions, and the progression of the model \cite{chhatwal_economic_2015}. First-order uncertainty is often introduced by the use of a random number generator through a deterministic program code; however, in digital computers random numbers are not really random. They are generated by an algorithm that produces a sequence of numbers that is seemingly random. These numbers are referred to as pseudo-random and the algorithm is called a pseudo-random number generator (PRNG). Each sequence produced by any PRNG is uniquely identified by its `seed' $s$, a number which provides the initial value to the generator. The seed is usually supplied by an environmental variable such as the computer clock, which can virtually guarantee that it is different for every simulation run. In other words, a PRNG produces numbers based on a deterministic formula which is seeded with some initial number. This allows for computer, and particularly agent-based models to simulate stochastic variables but also offer reproducible results. First-order uncertainty can be reduced by running the model several times, commonly known as `Monte Carlo' simulations \cite{ligmann2014,chhatwal_economic_2015}. 
The sufficient number of simulations usually depends on some descriptive statistic of the model output, typically the means and variances, where the the minimum number is the point when the statistic reaches some stability \cite{lee_complexities_2015}. While  suffering from some degree of subjectivity, \cite{lee_complexities_2015} this provides expositions on calculating the minimum number of simulations taking into account variance stability, effect size, and multivariate stability. In general, the sufficient number of Monte Carlo simulations depends on the parameters, time horizon, and the model structure, but there is no specific mathematical formulae to determine the minimum number of simulation runs. 

The behavior of any ABM is also sensitive to the model parameters and on the initial conditions, often referred to as `second-order uncertainty' \cite{chhatwal_economic_2015}. While first-order uncertainty relates to stochasticity in model structure, second-order uncertainty corresponds to estimation of parameters since true values are often unknown. In order to address second-order uncertainty, underlying distributions of parameters should be utilized whenever available. Second-order uncertainty can also be addressed by the use of sensitivity analysis methods such as one-parameter-at-a-time local, least squares linear fit, linear regression, Gaussian processes, and Latin Hypercube Sampling techniques \cite{mckay1979comparison,marino2008methodology,fonoberova20138}.

\subsection{Temporal Evolution}
\label{sec:TE}
Often, in ABM, the model components (e.g., environment) may not only react to the agent inputs but also evolve according to endogenous factors including time. Time evolution can be modelled in three ways: (i) `continuous time' in which the model can compute the system state for any time input, (ii) `discrete time' in which time evolves in distinct, but fixed intervals, and (iii) `discrete event' in which time instantaneously jumps from one event to the next. Within each time step, specific events may occur (e.g., movement of agents across the environment or transmission and progression of disease) or agents update their internal states based on interactions and endogenous factors. The decision to use a type of time evolution depends on the application domain, and most ABM computational systems implement discrete time or event transitions.   
\paragraph{Discrete time}
In a discrete time structure, the simulator engine advances the virtual clock by a given interval $\Delta t$. The time interval  $\Delta t$ will typically have a natural real-world unit associated with it, such as hours, days, or years depending on the phenomenon underlying the model.  For example, an agent-based model that tracks the infection dynamics of HIV over time may have the time units of `year', while a model for influenza infection is often run with the units of `day' resolution. At each $\Delta t$, the simulator engine picks agents sequentially (in some given order) and executes their PDA cycle iteratively. An inherent limitation to this method is the order in which the agents are selected (and their PDAs are executed) which may change the result obtained for the system, since the system state evolves with each action. It is entirely possible that a different ordering may result in vastly different global dynamics. 
Possible solutions to this limitation include selecting shuffling the order of agents at every time interval or have all agents simulate concurrently by operating on temporary variables, so that the perceived state is the same for all agents. 

\paragraph{Discrete event}
This simulation process considers time evolution as discrete increments with variable magnitudes corresponding to events occurring in the model. Every jump in time marks a change in the state of system. Between consecutive events, the system is idle and no changes occur. A discrete-event model is programmed to maintain a `queue' corresponding to the list of events that will occur. The program iterates over the queue, executing the event at the head of the queue. Once the event executed, it is removed from the queue, and the next event is scheduled appropriately (and sometimes dynamically). In comparison to the discrete-time approach, this is a more flexible paradigm. For instance, because discrete-event simulations do not have to simulate every time-step, they are typically run faster than the corresponding discrete-time simulations.

\subsection{Validation and Calibration}
\label{sec:validation}

Validation concerns with identifying the degree of consistency between an agent-based model and the underlying system it represents. It comprises of two stages: (i) `Input validation' that refers to the realism of the assumptions used to build the model, and (ii) `output validation' which measures the plausibility of the model outcomes relative to the observations of a real-world phenomenon. Output validation relies on a process called `calibration', which systematically refines the model parameters so that the output data closely resemble those observed in the phenomenon. A well-calibrated and validated model can be used to make predictions, and can provide inference on micro states, which may not be inferred from standard time series or statistical methods.

A proposed framework of validity \cite{marks2007ComputEcon} characterizes a model as \textit{useful} if it can exhibit at least some real-world observations; as \textit{accurate} if the simulated data matches historically observed real-world data; and as \textit{complete} if the simulated data matches all of the observed patterns of the real-world phenomenon. Based on this framework, the goal of validation and calibration is to construct a model that can be accurate, but also complete if possible.

\paragraph{Input validation} This stage validates the structure and assumptions of the model relative to the theory it is based on. Structural assumptions include choices of the rules and behaviours that define an agent's PDA cycle, the environment, and pattern of interactions. For instance, agents can be utility maximizing or employ bounded rationality. A more accurate and complete set of assumptions correspond to a higher number of parameters and variables in the model. Parameter-rich models are often difficult to validate, may suffer from over-fitting, and can end up in \textit{dimension hell} and might even become computationally infeasible.  In order to cope with these challenges in an accurate and complete model, it is often the case that input validation is conducted against some stylized facts (defined as a simplified presentation of an empirical finding), i.e., focusing on a limited number of variables which are most relevant to the complex system. For example, the basic reproduction number (commonly denoted by $\mathcal{R}_0$) in epidemiology is a summary measure succinctly describing the burden of a communicable disease. It defines the number of new cases that an infectious individual can generate in an entirely susceptible population \cite{diekmann2000mathematical}. In this sense, an agent-based model of a communicable disease may be validated by choosing parameters that yield the desired reproduction number from the simulated data. 

Input validation also often requires `program validation' which refers to the validation of the simulator engine. This aims to validate the computer codes which implement the various components of an agent-based model. Computer codes do not automatically generate errors when they encounter a bug or if some rules is incorrectly implemented. On the contrary, most programs will continue to produce results, independently of how the computational code is built. Careful attention is therefore needed to capture bugs and other artifacts. It is recommended that one should follow the principles from computer-science, including modularity and unit-testing \cite{laskowski2014Biomat2013}. These tests can be performed for each functional component of the system during the model development cycle, in which simple scenarios are created to verifying that all modules of the model are working in concert when executed together. In addition to unit-testing, well-established methods of degeneracy tests can always be performed to evaluate the outcomes using extreme value analysis by selectively disabling portions of the model or selecting plausible values for input parameters from estimated ranges \cite{sargent1987Proc19thConfWinterSimul-WSC87}.

\paragraph{Calibration} This stage systematically determines a set of parameters and input values that maximize the fitness of the model to the observed data, i.e., finding parameter values, assumptions and structural components that make the model `fit the data well' \cite{Vanni2011}. Calibration of agent-based models can be inherently difficult due to (often) a large parameter space, long simulation run-times, and stochasticity of events. The high dimensional nature of the model further exacerbates the issue of calibration.

The calibration process usually consists of running the model and evaluating a suitable summary measure of the simulated data, and comparing that against a set of empirically observed data such as time-series or experimental measurements \cite{Vanni2011}. For example, a common choice for such a measure is the squared difference
\begin{equation*}
 \text{d}(\text{data}_{\text{simulated}},  \text{data}_{\text{observed}}) = (\text{data}_{\text{simulated}} - \text{data}_{\text{observed}})^2,
\end{equation*}
which increasingly penalizes parameters that make the simulated data more distant from the empirical data. The summary measure, in general, is chosen with respect to the context of the model and can be, for example, specific data points, cross-sectional averages, regression analysis, and averages over realisations. If the simulated dynamics bear resemblance to the real-world observations, then the model is a possible explanation of the underlying complex system.

However, this way of calibration  may pose several challenges. First, it is often computationally demanding. Indeed, for any given point in the parameter space, a large number of Monte Carlo simulations may be needed to generate a distribution for the statistics of interest or by comparing instances of a model with different values of parameters and choosing the ones that best fit the data. Second, this calibration approach does not necessarily imply that the model reaches a single optimal choice for input parameters. Indeed, the reproduction of empirical data by a set of statistical properties of the simulated data is a quite weak form of calibration, since, in general there may be many possible sets of parameter values that generate the same outputs given the high degrees of freedom in the model. However, one can generate confidence intervals in which the true value of the parameters lie \cite{Vanni2011}. In a Bayesian approach to calibration, parameters are assigned a posterior probability distribution taking into account the uncertainty about their values based on prior knowledge. This effectively uses observed empirical data to inform parameters such that the model predicts the past and present observations well \cite{grazzini2017JEconDynControl}.  Lastly, it is clear that this calibration process is limited by the quality of the available data. When restrained by the lack or limited amount of quantitative data, qualitative data may be used for parameter identification \cite{mitra2018}. Nevertheless, the preceding two decades have seen an increasingly number of studies that attempted to calibration and validate agent-based models by way of optimization techniques and statistical methods, taking into account both quantitative and qualitative data \cite{richiardi2018Agent-basedModelsa,grazzini2015JEconDynControl,grazzini2017JEconDynControl,Vanni2011, mitra2018}. These include simulated minimum distance, Bayesian estimation, Markov Chain Monte Carlo, Sequential Monte Carlo, and particle filters methods. 

\section{A Mathematical Formalism of ABM}
\label{sec:AMFABM}
Considering the fundamentals of the ABM approach described above, we now provide an overview for its mathematical representation. 
\subsection{Partial Recursive Functions}
Since ABM is a collection of independent models, all of which are computer programs, there exists a corresponding unique partial recursive function  \cite{epstein2006HandbookofComputationalEconomics,richiardi2018Agent-basedModels}. Therefore, in principle, one can provide a representation of ABM as an explicit set of mathematical formulas (i.e., recursive functions). A system of equations is recursive if the output depends on one or more of its past outputs. In other words, the values for all state variables can be determined sequentially rather than simultaneously. If, in addition, the probability distribution of the next state depends only on the current state (and not any other previous states), the system is \textit{memoryless} and is called a `Markov Chain'. In fact, it is always possible to characterize an ABM as a Markov process \cite{IzquierdoJass} by redefining the state space. 
Thus, agent-based models can be created as Markov chains. 

To begin, consider an agent-based model $\mathcal{A}$ with $m$ agents which evolves over some time, and can be continuous or discrete. We are interested in the state of the model observed at discrete times $t_1, t_2, \ldots, t_T$, with $t_k < t_{k+1}$. Even if the underlying model runs in continuous time, the model state can be sampled at discrete observation times. At any time $t$, an agent $a_i \in \mathcal{A}$ is associated with the variable $\vbf{x}_{(i, t)}$, which take values in some finite field $\mathbb{F}$. Assuming the variables can be codified in a finite set of possibilities and are quantitative in nature (i.e., real numbers), then $\vbf{x}_{(i, t)} \in \mathbb{R}^n$ corresponds to an $n$-dimensional vector. The variable $\vbf{x}_{(i, t)}$ represents the state of an agent. For example, an agent could be described by the vector (age, sex, location), so the set of variables is a triple with a mixture of numerical and codified entries.
Borrowing notation from \cite{richiardi2018Agent-basedModelsb}, the evolution of the agent's state variables through time is specified by the equation:
\begin{equation}
\vbf{x}_{(i, t+1)} = f_i(\vbf{x}_{(i, t)}, \vbf{x}_{(i_-, t)}, \alpha_i),
\label{abm:eqn:updatefunction}
\end{equation}
where $f_i$ is an agent-specific update function (or transition function), which internally implements the agent's PDA cycle, $\alpha_i$ is a vector of agent-specific parameters (some of which could be stochastic), and $\vbf{x}_{(i_-, t)}$ is the state of all agents except $a_i$. The set of update functions $f_i$, one for each agent $a_i$, defines the `data-generating process' (DGP) of the model. These functions may be complicated, possibly involving discontinuities, fuzzy logic rules, and `if-else' statements.
In the case that each $f_i$ is a polynomial, the resulting model is called a polynomial dynamical system and is amenable to the computational tools and
theoretical results of computer algebra \cite{veliz-cuba2010Bioinformatics}.
The complete system state at time $t$ is the collection $\vbf{X}_{t} = [ \vbf{x}_{(1, t)} \; \vbf{x}_{(2, t)} \ldots \vbf{x}_{(m, t)} ]$ which is an $n \times m$ matrix of all individual states. The time evolution of the overall model is thus specified as 
\begin{equation}
\vbf{X}_{t+1} = F(\vbf{X}_{t}, \vbf{\alpha}) + \xi_{t},
\end{equation}
where $\vbf{\alpha}$ a vector of agent-specific parameters, and $\xi_t \in \mathbb{R}^{n \times m}$ is a matrix containing all stochastic elements at time $t$. Since the DGP functions $f_i$ may be nonlinear with stochasticity, the agent-based model is better represented by a more general map $\mathbf{F}$:
\begin{equation}
\label{abm:eqn:transitionequation}
\vbf{X}_{k+1} = \mathbf{F}(\vbf{X}_{k}, \vbf{\alpha}, \vbf{\xi}_{k}),
\end{equation}
where $\vbf{\xi}_{t}$ is a stochastic random matrix. The initial conditions of the system at $t_0$ are $(\vbf{X}_0, \vbf{\xi}_0)$. 
Equation~\eqref{abm:eqn:transitionequation} is called the \textit{transition equation} of the system.  A closer look at \eqref{abm:eqn:transitionequation} reveals the Markov chain representation of an agent-based model, though in practice these expressions may be complicated and difficult to interpret. 
Indeed, in most cases, the explicit set of functions $f_i$ and consequently $\vbf{F}$, are not mathematically tractable. In analytical models, the transition functions in \eqref{abm:eqn:transitionequation} often have closed form, with a simple structure, linear (or linearized), and are kept free (or at a minimum level) of heterogeneity. Any aggregation can be performed on variables by taking expectations over the stochastic elements. However, in agent-based models, the specification of \eqref{abm:eqn:transitionequation} have little or no restrictions. Since the state space of the model can grow large (possibly with infinite states), the transition equation may not have an analytical representation. 

Once the DGP has been specified, we are then interested in some aggregate or macro features of the model. Let $\vbf{y}_t$ be a set of aggregate statistics at time $t$, and let $h$ be a statistic function (i.e., a projection from $\vbf{X}$ to $\vbf{Y}$) over the system state, given by
\begin{equation}
\label{abm:eqn:aggregate}
\vbf{y}_t = h(\vbf{x}_{(1, t)}, \, \vbf{x}_{(2, t)}, \ldots, \vbf{x}_{(m, t)}) = h(\vbf{X}_t). 
\end{equation}
Regardless of the complexity and specification of each $f_i$, the solution (i.e., shape and form) to \eqref{abm:eqn:aggregate} at each time $t$ can always be found by backward iterations, which trace $\vbf{y}_k$ back to the initial conditions. That is, 
\begin{align*}
\vbf{y}_0 &=  h(\vbf{X}_0), \\
\vbf{y}_1 &=  h(\vbf{X}_1) = h(\vbf{F}(\vbf{X}_{0}, \vbf\alpha, \vbf{\xi}_{0})), \\            
\vbf{y}_2 &=  h(\vbf{X}_2) = h(\vbf{F}(\vbf{X}_{1},  \vbf\alpha, \vbf{\xi}_{1}) ) =  h( \vbf{F}(\vbf{F}(\vbf{X}_{0}, \alpha, \vbf{\xi}_{0}), \vbf\alpha, \vbf{\xi}_{1}) ), \\             
&\;\;\vdots \notag \\            
\vbf{y}_k &=  h( \vbf{F}(\ldots \vbf{F}( \vbf{F}(\vbf{X}_{0}, \vbf\alpha, \vbf{\xi}_{0}), \ldots) ) ).
\end{align*}
This backward iterations process uniquely relates the value of $\vbf{y}_t$ to the initial conditions, however explicating this relationship may be difficult because of the stochastic $\vbf\xi_t$ terms.  Since the DGP functions in Equation~\ref{abm:eqn:transitionequation} and the statistic function $h$ may be nonlinear, these random terms cannot be averaged out by expectations. Therefore, the relationship between the initial conditions $(\vbf{X}_0, \vbf{\xi}_0)$ and the statistic $\vbf{Y}$ is only realized by Monte Carlo analysis. Using Monte Carlo simulations for different initial states and parameter values, one could obtain a distribution for $\vbf{Y}$. Recall that Monte Carlo techniques rely on a pseudo-random number generator which is an inherently deterministic algorithm given the initial value of the seed $s$. Thus, any stochasticity implemented in the model by virtue of a PRNG has a deterministic nature, which allows us to further explore the formalism of an ABM. In particular, the stochastic term $\vbf{\xi}_k$ is a deterministic function of the seed $s$ and can be considered, conveniently, as part of the initial conditions. Letting $\vbf{Z}_0 = \{ \vbf{X_0}, s \}$, \eqref{abm:eqn:transitionequation} reduces to
\begin{equation}
\label{abm:eqn:transitionequation_reduced}
\vbf{X}_{k+1} = \vbf{F}(\vbf{Z_0}, \alpha) .
\end{equation}
It then follows that the statistic $\vbf{Y}$ is given by:
\begin{equation} 
\begin{split} 
\vbf{X}_{k} &= \vbf{F}( \vbf{F} ( \ldots \vbf{F}(X_0, \alpha, s))) = \vbf{F}^k(\vbf{Z}_0, \alpha) ,\\
\vbf{y}_k   &=  h( \vbf{F}^k(\vbf{Z}_0, \alpha) ) \equiv g_k(\vbf{Z}_0, \alpha). \label{abm:eqn:iot}
\end{split}
\end{equation}
Equation \eqref{abm:eqn:iot} is called the `input-output transformation' (IOT) function and drives the results of the ABM. When the seed $s$ is fixed, the IOT is a deterministic mapping of inputs (i.e., initial values and parameters of the system) onto outputs. While this is a convenient mathematical representation, from a practical point of view, the explicit form of the IOT function is unknown and the empirical distribution of the underlying stochastic process is often obtained by Monte Carlo simulations, by selecting different random seeds together with the initial conditions and parameter values. However, since an ABM places little restrictions on the specification of \eqref{abm:eqn:updatefunction}, careful attention is needed to control for the complexity, within the bounds of available computational resources. 
Nevertheless, we have provided a convenient formalism for agent-based modelling to bridge the gap between analytical models and computer simulations. This formalization is abstract enough to apply statistical rigour in performing quantitative analysis of the emergent properties of an agent-based model, in particular to assess stationarity and ergodicity. 

\subsection{Stationarity and Ergodicity}
In ABM, stationarity and ergodicity tests are crucial to know whether the model reaches a statistical (unique) equilibrium state \cite{grazzini2012JArtifSocSocSimul,richiardi2006}. These are intuitive concepts describing the long-term properties of a process or model. Stationarity of a process, in general, implies that every observation comes from the same probability distribution and that every observation carries information about the properties of the DGP. A variety of well-established parametric techniques exist for understanding stationarity of traditional models. In the framework of agent-based models where the implementation of the DGP may not yield an analytical form, one needs to confront with \textit{a priori} unknown stochastic properties of the model, assumptions and applicability of parametric tests that may be too restrictive or erroneous. Therefore, non-parametric tests are in general more suited for agent-based models as they do not require any assumptions on the IOT function of the model.

As mentioned before, the autonomous and heterogeneous nature of agents can evolve the system and generate the global dynamics. Equilibria associated with this global behaviour are idiosyncratic with respect to the agents, in the sense that in distinct Monte Carlo simulations, the evolution of every agent may vary substantially. Therefore, equilibria in agent-based models can only be defined at the aggregate level and in statistical terms after the global dynamics have emerged. When the model is relatively simple so that for any values of the parameters, it is stationary and ergodic, it is generally possible to characterise its equilibria. However, non-stationarity and non-ergodicity preclude the possibility of fully describing the long-term dynamics of the model. By stationarity in this section, we mean weak stationarity (also known as covariance stationarity).
\begin{defn}
	A stochastic process $\{w_t\}$ is weakly stationary if the first moment of $w_t$ is independent of $t$, that is, $\operatorname{E}(w_t) = \mu$ and if $\operatorname{Cov}[w_t,w_{t+h}]$ exists, is finite and depends only on $h$ and not on $t$. 
\end{defn}

Recall the IOT function $\{ \vbf{y}_k \}_{k = 0}^\infty$ defined by $\vbf{y}_k =  g_k(\vbf{Z}_0, \alpha)$, which relates the initial state of the system $Z_0 = (\vbf{X}_0, s)$ to the aggregate output of the model $\vbf{y}_k$. We call $\{ \vbf{y}_k \}_{k = 0}^\infty$ the associated time-series of the agent-based model. 
\begin{defn}
	\label{abm:defn:equilibrium}
	A statistical equilibrium in an agent-based model is reached in a given time window $(t^-, t^+)$ if the associated time-series $\{ \vbf{y}_k \}_{k = 0}^\infty$ is (weakly) stationary. The statistical equilibrium is denoted by $\mu^* = g^*(Z_0)$ and is given by 
	\begin{equation}
	\mu^*(Z_0, \alpha) = \operatorname{E}[y_t \mid t \in (t^-, t^+)],
	\end{equation}
	with respect to the process $\{ y_t \}$, and initial conditions $Z_0$. An equilibrium is said to be an absorbing (or steady-state) if $\vbf{y}_k$ is stationary in $(t^-, t^+ + \tau), \tau \rightarrow \infty$. An equilibrium is said to be a transient if $\vbf{y}_k$ is stationary in $(t^-, t^+)$, but no longer stationary in $(t^-, t^+ + \tau), \tau > 0$.
\end{defn}

A model may display both transient and absorbing equilibria, but the latter shows that once the system is in this state, it can no longer move out.  On the other hand, a model may oscillate between two or more transient equilibria (possibly followed by an absorbing equilibria). It follows that for any given initial conditions and parameter values, there can be at most one absorbing equilibrium. It is entirely possible that a model displays no absorbing equilibrium for a given statistic of interest. If a model is stationary and converges to the same equilibrium $\mu^*(Z_0, \alpha) = \mu^*(\alpha)$ irrespective of the initial conditions $\vbf{Z}_0$, the process $\vbf{y}_k$ is said to be `ergodic'. 
Ergodicity is sometimes defined \cite{grazzini2018Agent-basedModels} as: 
\begin{equation}
\label{abm:eqn:ergodic}
\lim_{n \rightarrow \infty} \frac{1}{n}\sum_{k=1}^n \operatorname{Cov}(y_t, y_{t-k}) = 0,
\end{equation}
which describes a property that concerns with the memory of a process. An ergodic process is characterized by weak memory (low persistence), and events far away from each other can be considered as almost independent because the effects of stochasticity fade with time. In other words, if an equilibrium is reached, it will be the same for all simulation runs, irrespective of the initial conditions, and the absorbing equilibrium will be \textit{unique}. Different initial values $(X_0, s)$ would not change the equilibrium value $\mu^*$, but might change the reaching time. Moreover, if $\vbf{y}_k$ is ergodic, the observation of a unique time series provides sufficient information to infer the shape and form of the IOT function in \eqref{abm:eqn:iot}. This means that if the model is ergodic, the properties can be analysed by using a time-series produced by a single run of the model. If the model is non-ergodic, then a set of Monte Carlo simulations (each produced by the same IOT but with different seeds) are necessary to describe, in distributional terms, the properties of the model.  

The inherent lack of an analytical form of the DGP and the difficulty of dealing with unknown stochasticity require non-parametric statistics and tests. A common standard non-parametric test used to check stationarity is an application of the `Runs Test' developed by Wald and Wolfowitz in 1940 \cite{grazzini2012JArtifSocSocSimul}, which tests the hypothesis that a given set of observations are mutually independent and randomly distributed.  While ergodicity is crucial for understanding the long-term behaviour, literature surveying tests for ergodicity is rather scarce. A previous work \cite{grazzini2012JArtifSocSocSimul} describes a modified Runs Test algorithm, which considers the invariance of the moment of order $k$ between different time-series produced by the same DGP, but with different random seeds. A full detailed survey of stationarity tests can be found in \cite{grazzini2012JArtifSocSocSimul,phillips1998JEconSurv}.

\section{Application of ABM to Disease Dynamics}
The use of ABM in the filed of mathematical epidemiology has been rapidly growing, with the development of comprehensive models that incorporate various databases to address public health challenges \cite{badham2018developing,najafi2017effect,shoukat2018BMCMed,shoukat2018cost}, in particular for emerging infectious diseases \cite{venkatramanan2018using,zhang2017spread,moghadas2017asymptomatic}. In this section, we detail an application of ABM to disease dynamics. We illustrate the construction and calibration of an agent-based model that describes the dynamics of disease transmission in a simple linear cascade of infection and recovery. The model we consider here was originally developed by Kermack and McKendrick in the 1920s \cite{m1925applications,kermack1927contribution}, and is referred to as the classical SIR (Susceptible-Infected-Recovered) model. In this model, the population is stratified into three different compartments (or health states) of susceptible ($S$), infected ($I$), and recovered ($R$).  A susceptible individual leaves the $S$-compartment when infected and enters the $I$-compartment. Similarly, an infected individual leaves the $I$-compartment and enters the $R$-compartment upon recovery. When the rate of infection is proportional to the total number of individuals in the population, the model can be represented by a set of differential equations:
\begin{equation}
\label{SIR}
\begin{split}
	\frac{dS}{dt} &=-\beta \frac{SI}{N}, \\
	\frac{dI}{dt} &= \beta \frac{SI}{N} - \gamma I, \\
	\frac{dR}{dt} &= \gamma I, 
\end{split}
\end{equation}
where $\beta$ is the rate of disease transmission, $\gamma$ is the recovery rate, and $N = S + I + R$. Although \eqref{SIR} is written in deterministic form, it is clear that a disease transmission process involves stochasticity as contacts between individuals occur randomly, even when stochastic nature of other behavioural, host, and biological factors are omitted. Thus, the classical SIR model is built on the assumption of  homogeneous mixing in the population where all individuals have equal chance to interact with others. This means that, on average, each infected individual generates $\beta S/N$ new infected individuals per unit time. 

Here, we develop an agent-based model to replicate the dynamics of the SIR model in a simple structure. The general framework of the model includes two main entities: (i) an \textit{in-silico} two-dimensional lattice environment and (ii) a set of unique agents situated (fixed) in the lattice. We set the size of the lattice to $20 \times 20$ resulting in an environment with a total of $400$ agents. Each agent is fully characterized by their health status as \textit{Susceptible}, \textit{Infected}, or \textit{Recovered} which are programmatically codified as integer values of $0=\textsc{SUS}$, $1= \textsc{INF}$, and $2=\textsc{REC}$, respectively. Therefore, an agent $a$ is fully described by its associated internal state variable $x_{a} \in \{0, 1, 2 \}$. 

Running the agent-based model simply amounts to instantiating a fully susceptible agent population, introducing an infected agent as the initial condition, and iteratively letting the agents interact by executing their associated  PDA cycles. The iterative process operates over a discrete time-step where the simulator engine advances the virtual clock by a single unit, in which the PDA cycle of each agent is carried out sequentially.  The perception stage of each agent determines all possible interactions of the agent, modelled through contacts with up to 8 random agents on the lattice that are situated in neighbouring cells. The perception function returns the number of infected contacts $k$ out of the eight random contacts for each agent at any time-step. In the decision stage, decision functions encapsulating the logic of the interactions are executed. If a susceptible agent meets an infected agent, successful disease transmission is determined using a rejection sampling-based (Bernoulli) trial where the chance of success is defined by a suitable probability distribution. Letting $x_{a, t}$ denote the internal state of an agent $a$ at time $t$, the one-step transition probability (in the Markov process) is given by:
\begin{equation}
	\Pr\big[x_{a, t+1} = \textsc{INF} \mid x_{a, t} = \textsc{SUS}\big] = 1 - (1 - b)^k,
	\label{abm:eqn:sir_onestep}
\end{equation}
where $k$ is the number of simultaneous contacts with infectious individuals (assuming that transmission events are independent per contact) and $b$ is the transmission probability. If the trial is successful, a susceptible individual becomes infected. In a similar way, the one-step transition from \textsc{INF} $\rightarrow$ \textsc{REC} is given by:
\begin{equation}
	\Pr\big[x_{a, t+1} = \textsc{REC} \mid x_{a, t} = \textsc{INF}\big] = \begin{cases}
		1 \quad \text{if}\,\, t > t_{\text{U}} \\
		0 \quad \text{otherwise}
	\end{cases}
\end{equation}
where $t_{\text{U}}$ is an empirically derived parameter representing the infectious period, which in this context is sampled from a Uniform distribution between 3 and 6 units of time. Finally the action stage updates the internal state of each agent as well as the global state of the model and broadcasts it to the entire lattice for the next iteration of the PDA cycle to continue. The computational implementation of the associated update function is described in Algorithm \ref{abm:alg:sir_code}.

\begin{algorithm}
	\begin{algorithmic}[1]
	\State {\textbf{Input}\quad \textbf{:}agent $a$, time $t$}
	\If {$x_{a, t} = \textsc{SUS}$}
		\State $n \leftarrow \text{Discrete Uniform}\,[1, 8]$  
		\State $k = 0\,;$
		\For {1 to $n$}
		\State	$\hat{x} \leftarrow AgentState(\text{Discrete Uniform}\,[1, 400])\,;$
			\If {$\hat{x} = \text{SUS}$}
			\State	$k = k + 1$ 
			\EndIf
			\State $P = 1 - (1 - b)^k\,;$
			\If {rand() $< P$}
			\State $a.t_\text{U} \leftarrow \text{Discrete Uniform}\,[3, 6]$
			\State	$x_{a, t} = \textsc{INF}$
			\EndIf
		\EndFor
	\EndIf
	\If {$x_{a, t} = \textsc{INF}$} 
		\If {$t > a.t_{\text{U}}$} 
			\State$x_{a, t} = \textsc{REC}$
		\EndIf  
	\EndIf
	\caption{Pseudocode implementation of an agent's PDA cycle and the associated update function.}
	\label{abm:alg:sir_code}
	\end{algorithmic}
\end{algorithm}

A key parameter in our model is the \textit{unknown}  transmission probability $b$.  Typically the value of $b$ is calibrated to some empirical observation of the underlying system such as the basic reproduction number (denoted by $\mathcal{R}_0$ as described in Section \ref{sec:validation}) or incidence  rate (i.e., new infections per unit time). By running  Monte Carlo simulations, the value of $b$ could be estimated to match, for example, $\mathcal{R}_0$ obtained from simulation data with a given $\mathcal{R}_0$. It is worth noting that this process can be computationally demanding depending on the complexity of the model. One must sweep through a parameter space (which could be arbitrarily large), running Monte Carlo simulations for each value. Reduction of the parameter space to a suitable subset requires an educated initial guess. 
	
In our example, we calibrated $b$ to yield $\mathcal{R}_0 = 1.6$, indicating that an infected person can infect 1.6 individuals (on average) over the course of infectious period. The calibration procedure requires an initial value of $b$, and counting the number of secondary cases caused by the initial infected agent. If the predicted $\mathcal{R}_0$ is not acceptable, the value of $b$ is changed accordingly, and the process is repeated.  Of course, multiple realisations are necessary for each value of $b$ to address the first- and second-order uncertainties. Using 500 Monte-Carlo simulations, the calibration process provided an estimated value of $b=0.047$ for which the average of realisations gives $\mathcal{R}_0 = 1.6$.

\begin{figure}[htb!]
	\centering
	\includegraphics[width=\textwidth]{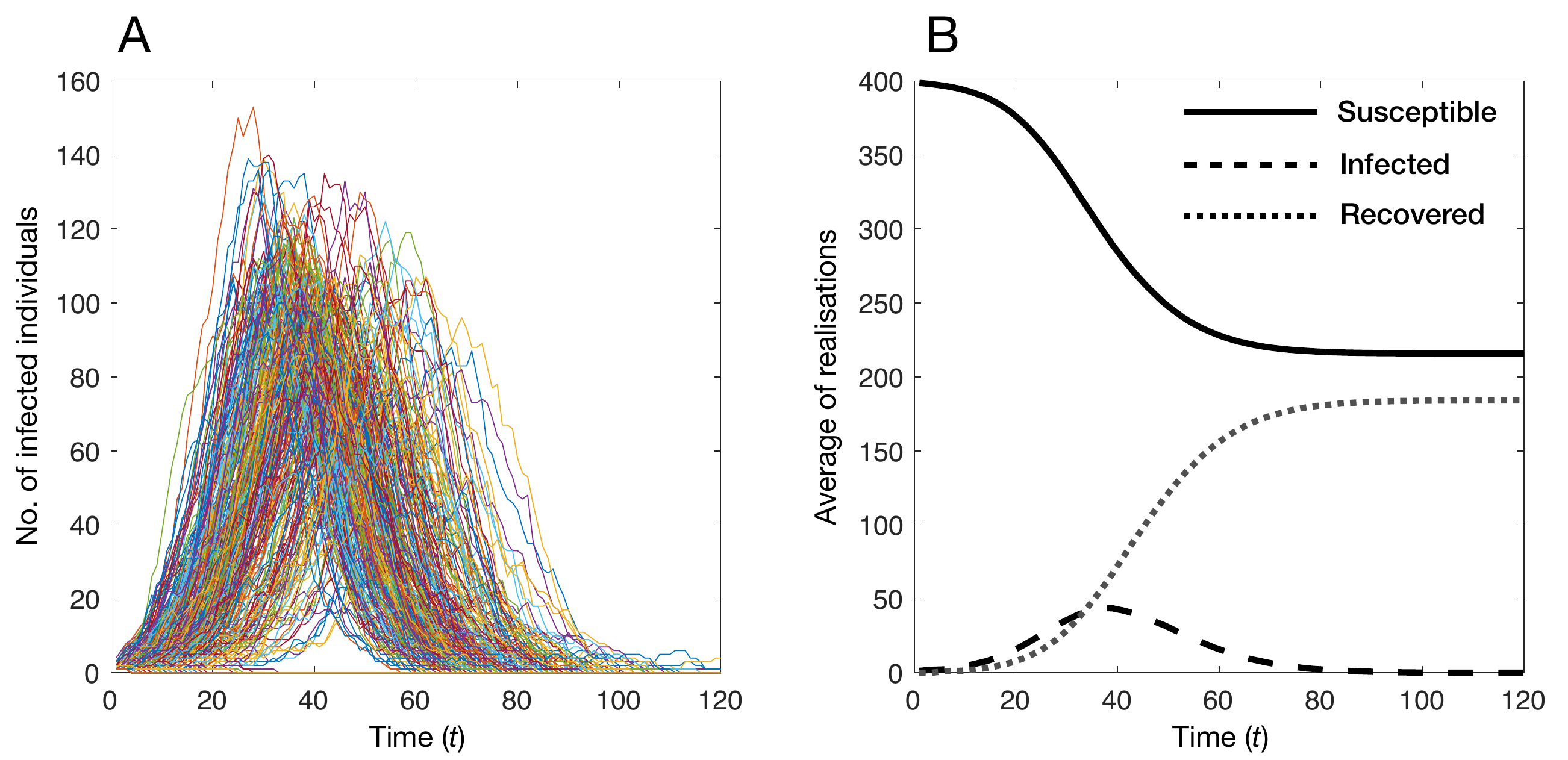}  
	\caption{Monte-Carlo simulations of the agent-based model for system \eqref{SIR}: (A) independent realisations of the state variable $I$; (B) average of realisations for all state variables.} 
	\label{fig:SIR}
\end{figure}

After calibration, we ran 500 Monte-Carlo simulation for 120 units of time to illustrate the behaviour of the system, corresponding to the spread of disease in the population. The global dynamics are represented by the changes in the number of individuals in different health states of the model. Figure \ref{fig:SIR}A shows the outputs for the state variable $I$ (i.e., the number of infections at any point in time) for each realisation. As is evident, each realisation produces a different infection curve as a result of stochasticity. It is also interesting to note that the average of realisations has a lower magnitude compared to many realisations. This is due to the fact that in many simulations, the initial infected case recovers without infecting any susceptible individuals and therefore the epidemic dies out. Figure \ref{fig:SIR}B shows the average of realisations for all state variables.

In the context of disease spread represented by the model in equations \eqref{SIR}, as the ABM follows agents through the environment, it is possible to provide more detailed (micro-level) information such as where an agent becomes infected or who infected them. This type of observables could lead to a better understanding of disease spread and can be useful to make decisions on control policies. However, simulating the model \eqref{SIR}, even when implemented stochastically, could not provide this agent-level information. Furthermore, recent studies show that a stochastic instantiation of equation-based models through implementation of a Markov process (e.g., Gillespie exact algorithm \cite{gillespie1977exact}) is still inadequate to produce some specific patters of the system behaviour observed in an ABM \cite{figueredo2014comparing}.

\section{Concluding Remarks}

The current surge of interest in ABM has gradually built up over the last twenty years \cite{bonabeau2002ProcNatlAcadSci}, especially with emerging technologies in computational power and big data collection platforms that bring a higher realism to such models for simulating the real-world phenomena. The use of agent-based models has provided an additional tool for advancing quantitative science, especially in research areas (e.g., public health domain \cite{tracy2018agent}) in which decision to intervene in the system dynamics may be subject to substantial heterogeneity and variability. While the capability of these models to address practical questions and inform decision-making in the face of uncertainty has been exemplified, there remain limitations to their systematic application. In particular, a more directed research is needed for expanding the theoretical aspects of ABM,  by taking into account the objectives of reliability, efficiency, and adaptability which underlie the flexibility of agent-based models. This remains a major undertaking, and our review here should only highlight its importance as the application of ABM expands in various disciplines. We hope that this exposition assists future research in this direction by providing a clearer understanding of the fundamentals and mathematical theories upon which ABM is established.

%\bibliography{references}
\printbibliography
% \noindent LaTeX formats citations and references automatically using the bibliography records in your .bib file, which you can edit via the project menu. Use the cite command for an inline citation, e.g.  \cite{Hao:gidmaps:2014}.

% For data citations of datasets uploaded to e.g. \emph{figshare}, please use the \verb|howpublished| option in the bib entry to specify the platform and the link, as in the \verb|Hao:gidmaps:2014| example in the sample bibliography file.

\section*{Acknowledgements}
%Acknowledgements should be brief, and should not include thanks to anonymous referees and editors, or effusive comments. Grant or contribution numbers may be acknowledged.
This study was in part supported by NSERC, Canada. 

%\section*{Author contributions statement}
% Must include all authors, identified by initials, for example:
% A.A. conceived the experiment(s),  A.A. and B.A. conducted the experiment(s), C.A. and D.A. analysed the results.  All authors reviewed the manuscript. 

% \section*{Additional information}

% To include, in this order: \textbf{Accession codes} (where applicable); \textbf{Competing interests} (mandatory statement). 

% The corresponding author is responsible for submitting a \href{http://www.nature.com/srep/policies/index.html#competing}{competing interests statement} on behalf of all authors of the paper. This statement must be included in the submitted article file.

\end{document}